**Increasing Market Penetration of LED Traffic Signals in New York State:**
**Review of Articles and Information on LED Traffic Signals**
*Lighting Research Center, Rensselaer Polytechnic Institute*
*American Council for an Energy Efficient Economy*
Prepared for the New York State Energy Research and Development Authority
***Final Draft:*** July 20, 2000

## ACKNOWLEDGMENTS

This review was conducted through the Lighting Transformations Program of the Lighting Research Center (LRC) at Rensselaer Polytechnic Institute and was supported by the New York State Energy Research and Development Authority (NYSERDA). Rachel Winters, NYSERDA Project Manager, oversaw the review. Mariana Figueiro is the Director of the Lighting Transformations Program. Suzanne Hayes of the LRC performed bibliographic searches for the materials described in this review. John Bullough of the LRC and Margaret Suozzo of the American Countil for an Energy Efficient Economy (ACEEE) prepared this summary.

## PURPOSE

A primary purpose of this review is to identify and summarize recently published information about the use and performance of light-emitting diode (LED) traffic signals. It consists of two sections: a synopsis of experiences of municipalities using LED traffic signals in 1999 and 2000, and a discussion of technical, economic and market issues surrounding the deployment of LED signals. This review includes articles and reports published in 1999 and 2000.

## EXPERIENCE WITH LED TRAFFIC SIGNALS

Table 1 summarizes articles that describe installations of LED traffic signals at various locations throughout the United States, Europe, Australia and New Zealand. This table treats the articles as responses to a hypothetical survey with the following questions:

- Where and when were LED traffic signals installed?
- How many LED signals were installed?
- What is the cost of LED signals?
- What financial assistance was available?
- Were energy savings experienced (or are they anticipated)?
- Were maintenance savings experiences (or are they anticipated)?
- What are the potential barriers to using LED signals?

Of the municipalities represented that described the LED signal colors being installed, 78% were installing red signals, 56% were installing green, and 11% were installing yellow signals (here and elsewhere in this summary, totals may add to more than 100% because some of the hypothetical questions above can have multiple answers). Of the municipalities installing green signals, 40% of them had previously installed red, or red and orange pedestrian signals. Half of the municipalities (50%) were installing LEDs in a significant number of their signals (10 or more intersections), and 30% were installing much smaller numbers of LED signals, often in just one or two intersections. These results seem to indicate a trend from previous years toward a greater use of green LED traffic signals and a greater penetration of LED signals overall.

There was a high recognition of the potential economic benefits of using LED traffic signals in the articles that were reviewed. Energy savings were cited by 80%, and maintenance savings by 90% of the municipalities. The ability to quantify (even if only crudely) these savings was demonstrated by 81% of the municipalities citing energy savings, and by 78% of those citing maintenance savings. This also seems to indicate a growing trend; typically, maintenance savings especially have been more difficult for municipalities to recognize and quantify (Conway and Bullough, 1999). Still, there was wide variation in reported "pay-back" periods, ranging from 1.5 years to more than 7 years.

Just under one third of the municipalities (30%) reported receiving some sort of financial assistance in their LED traffic signal programs. Half (50%) of these involved grants or other assistance from utilities, 50% involved assistance from other government agencies, and 17% was in the form of assistance from an energy service company.

More than one third (35%) of the municipalities cited potential barriers to more widespread use of LED traffic signals, and most of these (86%) were barriers associated with the very high initial cost of LED signals (in general, red LED signal heads were reported to cost around $100 and green heads about twice this amount). The only technological barriers cited (by 14% of those citing any potential barriers) were limitations of LEDs with respect to the yellow signals, which require much higher luminous intensities than red and green signals in North America.

**OTHER ISSUES SURROUNDING LED TRAFFIC SIGNAL USE**

*Visibility issues*
In June 1998, the Institute of Transportation Engineers (ITE) published its interim specification for LED traffic signal heads (ITE, 1998). The specification describes minimum luminous intensities of red, green and yellow signal heads and is based largely on the ITE's older existing specification for incandescent signals (ITE, 1985), but with a 15% across-the-board decrease in luminous intensity from the incandescent levels. In the ITE specifications, luminous intensities for green and yellow signals are higher than for the red signals. This differs from the approach taken by European nations and by Japan where all three signal colors are given the same luminous intensity. Recent research (Bullough *et al.*, 2000) indicates that for daytime viewing, signal visibility (e.g., response time and missed signals) is consistent with a requirement stipulating higher intensities for the green and yellow signals than for red signals.

The same study also indicates that for daytime viewing, there is no significant visibility difference between incandescent and LED traffic signals having the same nominal color and luminous intensity. The ITE will affirm or revise its interim standard in the coming months based on the results of a study funded by the National Cooperative Highway Research Program (NCHRP) and on other relevant research, perhaps including the results of Bullough *et al.* (2000). It has also been acknowledged that the faster onset times of LED signals compared to incandescent signals might be of benefit for signaling applications (Howe, 2000).

*Technical issues*
Key technical issues of concern to municipalities is how LED signals perform in actual installations. A number of technical challenges facing LED signals have been completely or partially overcome: these range from flicker to controller compatibility to lumen maintenance.

Most of the LED traffic signal market to date has consisted of retrofits designed to fit into existing incandescent signal lamp sockets. However, unique characteristics of LEDs may bring about opportunities for developments in new LED traffic signal systems. When viewed at night, for example, signals having very high intensity can result in discomfort or disability glare that impedes visibility. Feedback-controlled systems to reduce signal intensity at night or during overcast days have proven problematic with incandescent traffic signals because dimming an incandescent lamp also changes its color, and could create confusion to drivers about the color of a traffic signal. LEDs, on the other hand, do not significantly change in color when dimmed, and systems have been developed to take advantage of this potential (Johnstone, 1999). Another possible advantage of LED technology with respect to traffic control devices is the relative ease with which communication elements such as animation (Van Houten *et al.*, 1999), differing signal shapes (Anonymous, 1999n), and other options can be incorporated into signal designs (Pang *et al.*, 1999; Tam *et al.*, 1999). Such concepts, however, are not likely to find widespread use in many jurisdictions because traffic signals generally have highly standardized specifications; novel (and largely untested) configurations are not encouraged. Some municipalities, such as Philadelphia, PA (Suozzo, 2000), are encouraging new developments in LED traffic signal systems, such as three-color LED signals, through incentives to manufacturers.

Recent developments in LED traffic signal technology that could possibly encourage its more widespread use in the future include improvements in manufacturing processes for green and blue LEDs (Ranii, 1999; Reucroft and Swain, 2000) that can reduce production costs, and developments of durable plastic lens materials that are able to withstand large variations in temperature and severe levels of solar radiation (Anonymous, 2000d).

*Economic issues*
Certainly, municipalities have been challenged by the high initial cost of LED traffic signals and, in response, several electric utilities offer rebates or other financial incentives to municipalities for LED traffic signal installations. Declining prices are helping to overcome this barrier (Suozzo, 1999; Suozzo *et al.*, 2000). Several electric utilities have programs by which they offer rebates or other financial assistance for municipalities using LED traffic signals (CEE, 1999; Howe, 2000). Energy service companies, sometimes in partnership with LED signal manufacturers and financial institutions, have also begun working with municipalities to finance installation of LED signals with leases paid back through energy and maintenance savings (Anonymous, 1999o, 1999p). Such savings have been documented by organizations such as the Consortium for Energy Efficiency (CEE) and ACEEE (Suozzo, 1999; Anonymous, 2000e, 2000f). For example, the 7-year life cycle cost for incandescent versus LED red traffic signals is $288 for incandescent signals, compared to $154 for LED signals. Break-even points for green LED signals are estimated to be longer than for red LEDs (Anonymous, 2000g).

## SUMMARY

Use of LED traffic signals is growing, with greater penetration of not only red, but also green LED signals in greater proportion than previously reported. There appears to be growing acceptance of LEDs as viable light sources for traffic signals, and a growing awareness of the potential maintenance and energy savings achievable with LEDs, but specific tools to quantify such potential savings are still lacking. Providing such tools to agencies considering the use of LED traffic signals, and disseminating documented savings and long-term performance data, based on actual installations, would help reduce perceived risks (Suozzo *et al.*, 2000) associated with this technology.

In addition, there are a number of programmatic activities that have taken shape over the last couple of years, specifically the establishment of a national utility initiative by CEE to help coordinate utility promotion efforts, the development of a draft specification through the U.S. Environmental Protection Agency's and the U.S. Department of Energy's ENERGY STAR program (Suozzo *et al.*, 2000).

**Table 1.** Summary of articles and press releases documenting experiences of jurisdictions with LED traffic signals in 1999 and 2000.

| Location | Date installed | Quantity | Cost | Rebates/grants | Energy savings | Maintenance savings | Potential barriers |
|---|---|---|---|---|---|---|---|
| Anaheim, CA (Long, 1999) | 2000 | 273 intersections green (already have red) | - | assistance from Anaheim Public Utilities | 88%; $214,000/year | last 10 years | yellow technology not yet viable |
| Boston, MA (Anonymous, 1999e, 1999f; Palmer, 1999; Suozzo, 1999) | 1999 | 3600 red, 200 red arrows, 460 pedestrian signals in approximately 750 intersections | $425,000 with utility incentives; $100 per head | Boston Edison subsidizes price | 80% of the cost of operating red and orange signals (35% of total signal energy) | last 6 years versus 6 months; maintenance savings not yet quantified; anticipated reduction in emergency response | expensive |
| Christchurch, NZ (Anonymous, 1999a, 2000a; Neill, 1999) | 1999 | 1 signal | - | funding from Technology New Zealand | yes (would be $93,000/year total for all signals) | yes (would be $38,000/year total for all signals) | - |
| Denver, CO (Briggs, 2000) | 2000 | 13,000 green (already have red, orange) | $2.5 million total | - | 92%; $23/year per unit | $10-$12/year per unit | expensive |
| Elkhart, IN (Anonymous, 2000b) | 2000 | 2 intersections | - | - | - | - | - |
| Framingham, MA (Suozzo, 1999) | 1999 | 800+ signals: red, green, peds | $220,000 for signals; $25,000 for installation | Boston Edison incentives | More than 50% savings on energy bills | Number of calls requiring emergency response dropped dramatically after retrofit | - |
| Hamilton County, IN (Anonymous, 1999b) | 1999 | 10 intersections red | $100 per head | - | 14 versus 150 W, $8800/year savings | last 5-8 years | Yellow and green expensive |
| Kansas City, MO (Anonymous, 2000c) | 2000 | 1 intersection | - | - | - | yes | - |
| Kingston-upon-Hull, UK (Young, 1999) | 1999 | 26 signals | - | - | Yes | last 12 years versus 6 months | - |
| Lakewood, CA (Anonymous, 1999c) | 1999 | - | $150,970.72 total | state energy conservation grant | Yes | yes | - |
| Lee County, FL (Crawford, 1999) | 1999 | all red signals | - | - | - | yes | - |

| Location | Date installed | Quantity | Cost | Rebates/grants | Energy savings | Maintenance savings | Potential barriers |
|---|---|---|---|---|---|---|---|
| London and Bristol, UK (Hawkes and Nuttall, 1999; Pook *et al.*, 1999; Radford, 1999) | 1999 | 2 intersections | - | - | 80% | last 10 years versus 6 months, or 100,000 versus 1000 hours | - |
| Manchester, CT (Anonymous, 1999d) | 1999 | 54 red and 54 green | - | - | two-thirds less energy | yes | - |
| Nashua, NH (Anonymous, 1999g) | 1999 | - | - | - | 85%-90%; $36,000 in 2 years | last 7-10 years | Expensive |
| Newark, NJ (Anonymous, 1999h) | 1999 | - | - | - | yes | last 100,000 versus 5000 hours | - |
| Newton, MA (Suozzo 1999) | 1999 | 2222 signals red, green, peds | $627,000, materials and installation | Boston Edison incentives | more than 60% of previous energy bill | Not yet quantified | - |
| Nottingham, UK (Anonymous, 1999i) | 1999 | 1 intersection | - | - | one-third less energy use | last 10 years versus 6 months | expensive |
| Ocean Township, NJ (Anonymous, 1999j) | 1999 | 7 intersections red and green | $100 per head | $35 (8-inch) and $50 (12-inch) rebate from GPU Energy | $30,000/year total savings; 85% energy savings; last 7 years versus 12-14 months | | expensive |
| Overland Park, KS (Anonymous, 1999k) | 1999 | - | - | - | $160,000/year total savings | - | - |
| Sacramento County, CA (Anonymous, 1999l, 1999m) | 1999 | 118 intersections red | $250,000 total | $105,000 loan from California Energy Commission; $100,000 from Sacramento Municipal Utility District | 80%-90%; enough for 1.5 year payback considering energy alone | yes | - |
| Stockholm, Sweden (Jonsson, 1999) | 1999-2000 | 15,915 traffic and pedestrian heads (plus 11,085 already installed) | $3.2 million total; $461 for 3 heads | - | 6 versus 70 W; total savings of $470,900/year (for total 27,000 heads) | $243,000/year (for total 27,000 heads) | - |
| Victoria, Australia (Das, 1999) | 1999 | over 15-year period | $30 million total | - | - | - | - |
| Woburn, MA (Suozzo, 1999) | 1999 | 402 signals; red and green | - | Boston Edison incentives | Estimated to be 184,000 kWh per year at approx. $0.10 per kWh | Not tracked | - |